\documentclass[a4paper,12pt]{article}
\usepackage{amsfonts,slashed}
\usepackage{url}
\usepackage{latexsym}
\usepackage{amsfonts}
\usepackage{bm}
\usepackage{epsfig}
\usepackage{latexsym,amssymb}
\usepackage{amsmath,amssymb,amsthm}

\numberwithin{equation}{section}
\def\be{\begin{equation}}
\def\ee{\end{equation}}
\def\bea{\begin{eqnarray}}
\def\eea{\end{eqnarray}}
\def\ba{\begin{array}}
\def\ea{\end{array}}
\DeclareFontFamily{U}{bbold}{}
\DeclareFontShape{U}{bbold}{m}{n}
   {  <5> <6> <7> <8> <9> gen * bbold
      <10> <10.95> bbold10
      <12> <14.4> bbold12
      <17.28> <20.74> <24.88> bbold17
   }{}
\DeclareMathAlphabet{\mathbbold}{U}{bbold}{m}{n}

\newcommand{\M}{{\mathcal{M}}}
\newcommand{\LL}{{\mathcal{L}}}
\newcommand{\Tr}{{\rm Tr}}

\newcommand{\FF}{{\bf \mathfrak{F}}}
\newcommand{\GG}{{\bf \mathfrak{G}}}

\newcommand{\RR}{\mathbb{R}}
\newcommand{\CC}{\mathbb{C}}

\newcommand{\II}{\mathbbold{1}}
\newcommand{\OO}{\mathbbold{0}}
\newcommand{\nn}{\nonumber}
\newcommand{\no}{\nonumber \\}

\newcommand{\bbox}{\lower.2ex\hbox{$\Box$}}
\newcommand{\SU}{\mathop{\rm SU}}
\newcommand{\SO}{\mathop{\rm SO}}
\newcommand{\U}{\mathop{\rm {}U}}

\newcommand{\Symp}{\mathop{\rm {}Sp}}
\newcommand{\SL}{\mathop{\rm {}SL} }
\newcommand{\GL}{\mathop{\rm {}GL}}

\def\bfone{\relax{\rm 1\kern-.35em 1}}
\def\bfzero{\relax{\rm I\kern-.18em 0}}
\begin{document}
\vskip 8mm
\begin{center}
{\bf\LARGE  On Multifield Born and Born-Infeld Theories}\\\vskip 0.5cm {\bf\LARGE and their non-Abelian Generalizations} \\
\vskip 2 cm
{\bf \large Bianca L. Cerchiai$^{1,2,3}$ and Mario Trigiante$^{2,3}$}
\vskip 8mm
 \end{center}
\noindent {\small $^{1}$ Museo Storico della Fisica e Centro Studi e Ricerche Enrico Fermi, P.zza del Viminale 1, I-00184 Roma, Italia\\
$^{2}$ DISAT, Politecnico di Torino, Corso Duca
    degli Abruzzi 24, I-10129 Torino, Italia\\
    $^{3}$  Istituto Nazionale di
    Fisica Nucleare (INFN) Sezione di Torino, Italia}
\vskip 1.5 cm
\begin{center}
{\small {\bf Abstract}}
\end{center}
Starting from a recently proposed linear formulation in terms of auxiliary fields, we study $n$-field
generalizations of Born and Born-Infeld theories. In this description  the Lagrangian is quadratic in the vector field strengths
and the symmetry properties (including the characteristic self-duality) of the corresponding non-linear theory are manifest as on-shell duality symmetries and depend on the choice of the (homogeneous) manifold spanned by the auxiliary scalar fields and the symplectic frame. By suitably choosing these defining properties of the quadratic Lagrangian, we are able to reproduce some known multi-field Born-Infeld theories and to derive new non-linear models, such as the $n$-field Born theory. We also discuss non-Abelian generalizations of these theories obtained by choosing the vector fields in the adjoint representation of an off-shell compact global symmetry group $K$ and replacing them by non-Abelian, $K$-covariant field strengths, thus promoting $K$ to a gauge group.
\vskip 1 cm
\vfill
\noindent {\small{\it
    E-mail:  \\
{\tt bianca.cerchiai@polito.it}; \\
{\tt mario.trigiante@polito.it}}}
   \eject
   \numberwithin{equation}{section}
\section{Introduction}

The Born-Infeld Lagrangian in 4 dimensions~\cite{Born-Infeld}:
\be
\mathcal{L}=\mu^2 \left[1-\sqrt{\left| {\rm{Det}} \left( \eta_{\mu\nu}+{1 \over \mu}F_{\mu \nu}\right)\right|}\right]
=
\mu^2 \left[1-\sqrt{1+\frac{1}{2\mu^2} F_{\mu \nu} F^{\mu \nu}-
\frac{1}{16 \mu^4} \left(F_{\mu \nu} {}^* F {}^{\mu \nu}\right)^2}\, \right],
\label{1fieldBorn-Infeld}
\ee
where ${1 \over \mu}$ is a small real parameter, $\eta_{\mu \nu}$ the Minkowski metric, $F_{\mu \nu}=\partial_\mu A_\nu-\partial_\nu A_\mu$ an Abelian field strength and
${}^* F {}^{\mu \nu} $ its Hodge dual, was found as a generalization of electromagnetism in an effort to solve the problems posed by the self-interaction of the electromagnetic field generated by a point charge. Its main feature is that it is a non-linear Lagrangian, which is self-dual and implements electromagnetic duality in an interacting system.

Originally, Born~\cite{Born} had proposed a different Lagrangian:
\be
\mathcal{L}=\mu^2 \left[1-\sqrt{1+\frac{1}{2\mu^2} F_{\mu \nu} F^{\mu \nu}}\, \right]
=\mu^2 \left[1-\sqrt{\left| {\rm{Det}} \left( \eta_{\mu\nu}+{1 \over \mu}F_{\mu \nu}\right)\right|-\left| {\rm Det \left({1 \over \mu}F_{\mu \nu}\right)} \right|}\right].
\label{1fieldBorn}
\ee
The two Lagrangians describe the same static solutions~\cite{Born-Infeld,Born,Dirac}, and in general coincide if $\vec B=\vec 0$.  
They are both Poincar\'e invariant and self-dual. Notice that here by self-duality we mean the discrete $\mathbb{Z}_2$ symmetry under a Legendre transformation of all the fields~\cite{ZuIv, AFZ}. In the case of the Born-Infeld
theory this symmetry is enhanced to a continuous $U(1)$ electric-magnetic duality
~\cite{Sch, GR} analogous to the one of the Maxwell theory.

It was shown in~\cite{Born-Infeld, Born} that both Lagrangians predict for the electromagnetic field of a charged electron a finite energy and a non singular electric field.
The constant $\mu$ was identified with ${e \over 4\pi r_0^2}$, where $e$ is the charge and $r_0$ the classical electron radius.

The propagation of shock waves in non-linear electrodynamics \cite{Ple, MBCS} in general presents the phenomenon of birefringece in the vacuum, where the shock wave has two polarization modes traveling along distinct directions following distinct ``optical metrics". The main difference between the Born-Infeld and the Born theories is that while the latter does exhibit birefringence, the former model does not~\cite{MBCS}.

In string theory, Born-Infeld Lagrangian is relevant for the dynamics of open strings in a constant electromagnetic background~\cite{FT}. In particular, it is related to the effective action of D-branes~\cite{Tse1999}.

Moreover, the original Born-Infeld theory admits a supersymmetric extension and it turns out that when gauginos are added, there is a second hidden non-linearly realized supersymmetry~\cite{DP,CF}. In other words, the supersymmetric version represents the invariant action of the Goldstone multiplet in a $\mathcal{N} = 2$ supersymmetric theory spontaneously broken to $\mathcal{N} = 1$, with $\mu$ determining the supersymmetry breaking scale~\cite{BG,APT,FPS,Andrianopoli:2015rpa}.

Recently considerable effort has been devoted to multifield generalizations of the original Born-Infeld theory, which include not just vector fields but also antisymmetric tensors and scalars \cite{FAY2015, FAY2016}.
We shall restrict our analysis to four-dimensional models. A relation between some of these formulations is provided by the recently studied \emph{c-map} for Born-Infeld-like theories \cite{ADFT}. Multi-vector extensions of the Born-Infeld model, as well as the Born one, are most conveniently studied using the auxiliary field description developed in \cite{RoTs,ABMZ,ADAT}. This framework has the advantage that the global symmetries and, in particular, the distinctive self-duality of the Lagrangian, are built-in and manifest. For this reason it is also particularly suitable for discussing non-Abelian generalizations of Born-Infeld and Born-like multi-vector theories, which is one of the purposes of the present work. Indeed a characteristic feature of these non-linear theories is their off-shell global symmetry group $H_{e}$ (i.e. global symmetry of the Lagrangian), which originates from the corresponding symmetry of the quadratic action in the auxiliary field description and is one of its defining data\footnote{In fact it depends on the on-shell global symmetry group $H$ of the linear field equations and on the symplectic frame.}. If the vector fields are chosen to transform in the co-adjoint representation of a subgroup $K$ of $H_e$, the non-linear theory resulting from the integration of the non-dynamical scalar fields, can be made non-Abelian by simply replacing the Abelian field strengths by their non-Abelian counterparts:
\be
\partial_\mu A^I_\nu-\partial_\nu A^I_\mu\rightarrow \partial_\mu A^I_\nu-\partial_\nu A^I_\mu-g_c\,\,{{f}_{JK}}^I A^J_\mu A^K_\nu,\label{KcovF}
\ee
${{f}_{JK}}^I$ being the structure constants of $K$ and $g_c$ the coupling constant.
\par
The paper is organized as follows. First, in Section~\ref{linear} the auxiliary field description of multi-field extensions of Born and Born-Infeld theories, as outlined in~\cite{ADAT}, is recalled, including an analysis of the resulting symmetries of the non-linear theory. In addition, we propose a generalization to non-Abelian field strengths along the lines mentioned above. Subsequently, in Section~\ref{un} this formalism is compared with the Lagrangian introduced in~\cite{ABMZ}, in particular for the case of $\U(n)$-duality. Next, in Section~\ref{2field}, the method is applied to perform an explicit computation of 2-field Born-Infeld theories with $\U(1), \U(1) \times \U(1)$ and $\SU(2)$ symmetries, which are matched to the models previously obtained in~\cite{FAY2015, FAY2016}. Then, in Section~\ref{SOn} we present a new non-linear theory generalizing Born model to $n$ fields with $\SO(n)$ symmetry, that is self-dual and can be extended to a non-Abelian gauge symmetry. In Section~\ref{euler} we discuss our models in a different parametrization of the coset representative, generalizing the Euler angles. This coordinate frame turns out to dramatically simplify the expression for the potential, by making the symmetries manifest. Moreover the integration of the Euler coordinates yields, besides the non-linear model, also the Maxwell theory, which is self-dual as well.\par Finally, in Section~\ref{t3} we apply our technique to the manifold associated with the $t^3$-model in supergravity, and we study a consistent truncation for which the corresponding equations of motion can be explicitly solved.

\section{Auxiliary field formulation of Born and Born-Infeld Theories}
\label{linear}

We start recalling the main facts about the linear description of Born-Infeld and Born theories in terms of auxiliary fields, in the form worked out in~\cite{ADAT}:

\begin{equation}
\mathcal{L}=  -\frac{1}{4}\, {F}^T_{ {\mu} {\nu}}\,g\, {F}^{ {\mu} {\nu}}+
\frac{1}{4}\, {F}^T_{ {\mu} {\nu}}\,\theta\,{}^* {F}^{ {\mu} {\nu}}-\frac{\mu^2}{2}\,{\rm Tr}(N\mathcal{ M}) +\rm{const}.
\label{linbi}
\end{equation}

Here ${1 \over \mu}$ is a real parameter which plays the role of a perturbation parameter and should be taken small in order to obtain a well-defined non-linear description. $N$ is a constant $2n\times 2n$ real symmetric matrix,  which for the purposes of the current investigation can be set to the identity matrix.

The above Lagrangian describes the dynamics of the Abelian field-strengths $F_{\mu \nu}=\partial_\mu A_\nu-\partial_\nu A_\mu$ and of their Hodge duals ${}^* F^{\mu \nu}={1 \over 2} \varepsilon^{\mu \nu \lambda \sigma} F_{\lambda \sigma}$.

The auxiliary fields $g$ and $\theta$ are $n\times n$ real symmetric matrices, which depend on a set of scalars $\{\phi^s\}$. We assume the latter to be parameters of a homogeneous symmetric scalar manifold $G \over H$, admitting, in analogy with extended supergravity models, a flat symplectic structure \cite{GZ}, characterized by a symplectic, symmetric matrix $\M(\phi^s)$ of the form:
\begin{equation}
\mathcal{M}[g(\phi^s),\theta(\phi^s)]=\left(
\begin{array}{cc}
g+\theta g^{-1} \theta & -\theta g^{-1}\\
- g^{-1} \theta  &g^{-1}
\end{array}
\right)\in \Symp(2n),
\label{Mmatrix}
\end{equation}
 This matrix is the same which encodes the scalar couplings to the gauge field-strengths in extended supergravity theories and implements the embedding:
\be
\M: \frac{G}{H} \hookrightarrow {\Symp(2n) \over \U(n)}.
\ee
 The scalar fields lack of a kinetic term in the action so that they play the role of auxiliary fields with purely algebraic equations of motion.

Introducing the short-hand notation:
\be
\FF={F}^{{\mu} {\nu}}\, {F}^T_{ {\mu} {\nu}}, \quad {}^* \FF={F}^{{\mu} {\nu}}\, {}^* {F}^T_{ {\mu} {\nu}},
\ee
the Lagrangian (\ref{linbi}) can be conveniently rewritten as:
\be
\mathcal{L}=-\frac{1}{4} \Tr (\FF g) +\frac{1}{4}\Tr \left({}^*\FF \theta \right)-{\mu ^2 \over 2} \Tr\left(g+g^{-1}(\II+\theta^2)\right)+\rm{const},
\label{elinbi}
\ee
where we have set $N=\II$ and used the explicit expression (\ref{Mmatrix}) for $\M$ as well as the cyclic property of the trace for the potential.

Integrating out the non-dynamical scalar (auxiliary) sector $\{g(\phi^s)$, $\theta(\phi^s)\}$ through its equations of motion yields a non-linear $n$-vector Lagrangian.

In the one vector field case, choosing ${G \over H}={\Symp(2) \over \U(1)}$, the original non-linear Born-Infeld Lagrangian (\ref{1fieldBorn-Infeld}) is recovered \cite{ADAT}, while for ${G \over H}=O(1,1) \subset {\Symp(2) \over \U(1)}$, where:
\be
\M=\left(
\begin{array}{cc}
e^\phi&0\\
0&e^{-\phi}
\end{array}
\right),
\ee
we easily recover Born model~(\ref{1fieldBorn}).

Notice that the equations of motion for the scalar fields  are purely algebraic. By virtue of the flat symplectic structure defined over ${G \over H}$, the on-shell symmetry $H$ of the theory can be made manifest, according to the Gaillard-Zumino mechanism~\cite{GZ}. To this end let us
introduce the symplectic vector $\mathbb{F}=(F^I,G_J)$ composed by the (electric) field strengths $F^I_{\mu \nu}$ and their (magnetic) duals $${G_J}_{\mu\nu}=-\varepsilon_{\mu \nu\rho\sigma} {\delta \LL \over \delta F^J_{\rho \sigma}}\,.$$ The algebraic equations of motion for the scalar fields acquire the following manifestly symplectic-covariant form:
\be
\mathbb{F}^T \partial_s \M \mathbb{F}=-4\mu^2 \partial_s \Tr(\M).
\label{Heom}
\ee
On the other hand the field equations for $\mathbb{F},{}^* \mathbb{F}$ yield:
\be
\partial_{[\mu} \mathbb{F}_{\nu \rho]}=0,
\ee
and the ``twisted self-duality condition''~\cite{CJ}:
\be
{}^* \mathbb{F}=-\CC \M \mathbb{F}\,,
\label{twisted}
\ee
where:
\be
\CC=\left(
\begin{array}{cc}
\OO&\II\\
-\II&\OO
\end{array}
\right)
\ee
is the symplectic-invariant matrix. Let $k_\alpha^s$ generate infinitesimal isometry transformations in $G$:
$$\phi^s \rightarrow \phi^s+\epsilon^\alpha k_\alpha^s\,.$$
The symplectic structure on  ${G \over H}$ associates with each $k_\alpha^s$ a symplectic generator $t_\alpha$:
\begin{equation}
k_\alpha^s\,\,\rightarrow\,\,\,\,\,t_\alpha\in\mathfrak{sp}(2n)\,\,:\,\,\,\,\,t_\alpha \CC=-\CC\,t_\alpha^T\,,
\end{equation}
so that the transformation of the matrix $\M$ under an infinitesimal isometry generated by Killing vector $k_\alpha$ reads:
\be
\delta \M=\epsilon^\alpha k_\alpha^s \partial_s \M=\epsilon^\alpha(t_\alpha \M+\M t_\alpha^T).
\label{Mvariation}
\ee
The on-shell global invariance of the non-linear theory is given by
transformations $\phi^s\rightarrow \phi^{\prime s}$ which leave the potential part in (\ref{linbi}) unaltered:
\be
\Tr(\M(\phi^s))=\Tr(\M(\phi^{\prime s})).
\ee
This is the case if the transformation is associated with a symplectic matrix $A$ such that $\M(\phi^{\prime s})=A\M(\phi^s)A^T$ and $AA^T=\II$, i.e. $A\in {\rm Sp}(2n)\bigcap {\rm SO}(2n)={\rm U}(n)$. These conditions are satisfied by any transformation in $H=G\bigcap {\rm U}(n)\subset G$.
The invariance of the algebraic scalar field equations require, for any generator $t_\alpha$ of $H$:
\be
\mathbb{F}^T_{\mu\nu} t_\alpha \M \mathbb{F}^{\mu \nu}=0\,\,\Leftrightarrow\,\,\,\,\,\mathbb{F}^T_{\mu\nu}  t_\alpha\mathbb{C}{}^*F^{\mu \nu}=0\,,\label{GaZu}
\ee
where in deriving the last equations, also known as the Gaillard-Zumino conditions
~\cite{GZ},  we have used Eq. (\ref{twisted}). Note that a symmetry $A$ does not need to correspond to a non-trivial isometry of the scalar manifold. Indeed it can be an orthogonal, symplectic matrix commuting with $\M(\phi^s)$. In this case $ \phi^{\prime s}=\phi^s$. An example of such $A$ is the electric ${\rm U}(1)$ symmetry of the model to be discussed in Sect. \ref{U1U1}.

It is also interesting to notice that the invariance also contains the $\mathbb{Z}_2$ subgroup generated by the symplectic matrix $\CC$, which belongs to $\U(n) \subset \Symp(2n)$, but not necessarily to $H$. This discrete symmetry corresponds to the self-duality of the non-linear action obtained upon integration of the scalar fields. As a consequence, theories defined with this method are by construction self-dual.

Another important remark can be made about non-Abelian generalizations. As pointed out in the Introduction, if $H_e \subset H$ is the  global symmetry group of the action and $K \subset H_e$ a compact subgroup with respect to which $\{ A^I_\mu \}$ transform in the co-adjoint representation, the Born-Infeld and the Born theories can be promoted to non-Abelian, non-linear gauge theories with gauge group $K$ provided the field strength are $K$-covariantized as in (\ref{KcovF}).

In this respect, let us recall that for any compact group $K$ of dimension $n$, its adjoint representation Adj$(K)$ can be embedded \cite{Dyn} in the fundamental $n-$dimensional (vector) representation of $\SO(n)$.

Then, as an example, the $n-$field Lagrangian generalizing Born theory with duality symmetry $\SO(n)$, which arises from the embedding:
\be
{\GL(n)\over \SO(n)} \hookrightarrow {\Symp(2\, n,\RR) \over \U(n)},
\ee
can be gauged with any gauge group $K$ of dimension $n$. We study this Lagrangian in Sec.~\ref{SOn}, see Eq.~(\ref{LagSOn}). An analogous example is the $n-$field Lagrangian generalizing Born-Infeld theory with duality symmetry $\SO(n)$, obtained from the embedding
$
{\SL(n)\over \SO(n)} \hookrightarrow {\Symp(2\,n,\RR) \over \U(n)}.
$
Since, in the present work, we do not wish to make contact with string theory and D-brane actions, we shall not consider the symmetrization prescription used in \cite{Tse1997} in the definition of the non-Abelian Born-Infeld Lagrangian.  Our only starting point in constructing a non-Abelian Born or Born-Infeld-like theory is a multi-field, self-dual non-linear model.

\section{Born-Infeld Theories with ${\bm \U(n)}$ symmetry}
\label{un}
In this Section we compare the linear Lagrangian (\ref{linbi}) with the Lagrangian constructed in \cite{ABMZ}, proving that the model defined by the former is a consistent truncation of that defined by the latter. In \cite{ABMZ} the following Lagrangian is considered:
\be
\LL=\mbox{Re}  \, \Tr \left[ i (\lambda -S) \chi-{i \over 2}\lambda \chi S_2 \chi^\dagger+i \lambda ({F}^{\mu \nu} \, {\bar F}^T_{ {\mu} {\nu}}-i F^{ {\mu} {\nu}}\,{}^* {\bar F}^T_{ {\mu} {\nu}})\right].
\label{LagSABMZ}
\ee
Here $\lambda$ and $\chi$ play the role of the auxiliary fields, which are arbitrary complex $n-$dimensional matrices:
\be
\lambda=\lambda_1+i \lambda_2; \quad \chi=\chi_1+i \chi_2,
\ee
that are decomposed in terms of their real and imaginary parts with $\lambda_i, \chi_i$, $i=1,2$ hermitian matrices. For our purposes we set the extra auxiliary field $S=S_1+i S_2$, which would be necessary to implement the full $\Symp(2n)$ group, to $S=i$, i.e $S_1=0$, $S_2=i$. Then the Lagrangian (\ref{LagSABMZ}) reduces to:
\be
\LL={1 \over 2}\, \Tr \left[ i (\lambda -i) \chi-{i \over 2}\lambda \chi \chi^\dagger+i \lambda ({F}^{ {\mu} {\nu}} \, {\bar F}^T_{ {\mu} {\nu}}-i F_{ {\mu} {\nu}}\,{}^* {\bar F}^T_{ {\mu} {\nu}})+ \mbox{hermitian conjugate}\right].
\label{LagABMZ}
\ee

We integrate out $\chi$ by means of its equation of motion:
\be
{\delta \LL \over \delta \chi}=\lambda-i -i\chi^\dagger \lambda_2=0,
\ee
leading to:
\be
\chi=-\lambda_2^{-1}(1-i \lambda^\dagger); \quad \chi^\dagger=-(1+i \lambda) \lambda_2^{-1}.
\ee
Subsequently, we substitute these expressions for $\chi, \chi^\dagger$ in the potential:
\bea
\mathcal{V}&=&{1 \over 2} \, \Tr \left[ i (\lambda -i) \chi-{i \over 2}\lambda \chi \chi^\dagger +\mbox{h.c.} \right]\no
&=& {1 \over 2} \, \Tr \left[ -(1-i \lambda^\dagger)(1+i \lambda) \lambda_2^{-1}-{i \over 2} \lambda_2^{-1} \lambda \lambda_2^{-1}(1-i \lambda^\dagger)(1+i \lambda)+\mbox{h.c.}\right]\no
&=&{1 \over 2} \, \Tr \left[ -\left(\lambda_2^{-1}+{i\over 2} \lambda_2^{-1} \lambda \lambda_2^{-1}\right)(1-i \lambda^\dagger)(1+i \lambda)+\mbox{h.c.} \right]\no
&=&\Tr \left[ -\lambda_2^{-1}\left(1+\lambda^\dagger \lambda +i(\lambda-\lambda^\dagger)\right)\right]
\label{potABMZ} \\
&=& \Tr \left[ -\lambda_2^{-1}\left(1+\lambda_1^2+\lambda_2^2+i(\lambda_1\lambda_2
-\lambda_2 \lambda_1)\right)-2\right]\no
&=&\Tr \left[-\lambda_2^{-1} -\lambda_1^2\lambda_2^{-1}-\lambda_2\right]-2 \, \Tr (\II)\nonumber
\eea
We have applied the cyclicity of the trace and the fact that the antisymmetric parts of the expressions do not contribute.

With the identification:
\be
\lambda_2=g \mbox{ and } \lambda_1=\theta,
\label{ident}
\ee
the potential (\ref{potABMZ}) can be recast in the form:
\be
\mathcal{V}=-\Tr(\M)-2 \, \Tr(\II)\,,
\ee
where $\M$ was defined in (\ref{Mmatrix}) in terms of $g$ and $\theta$, which now, in this more general setting, are complex hermitian matrices.
The identification (\ref{ident}) is consistent with the fact that at this point, after the integration of $\chi, \chi^\dagger$, the Lagrangian (\ref{LagABMZ}) reads:
\be
\LL=-\Tr(\lambda_2 {F}^{ {\mu} {\nu}} \, {\bar F}^T_{\mu \nu})+\Tr(\lambda_1 F^{ {\mu} {\nu}}\,{}^* {\bar F}^T_{\mu \nu})-\Tr(\M)+\mbox{const}\,,
\label{LagABMZtr}
\ee
which, aside for the coefficients, has the same form as (\ref{elinbi}), except for the fact that $F^I$ are complex.\par Restricting to real $F^I$, in the first two traces on the right hand side of (\ref{LagABMZtr}), only the symmetric parts of $\lambda_1$ and $\lambda_2$ contribute. In the Lagrangian we can consistently truncate out the antisymmetric parts of $\lambda_1, \lambda_2$ so as to obtain (\ref{linbi}). Thus, when truncated from complex to real fields, the Lagrangian (\ref{LagABMZtr}) coincides with (\ref{linbi}), once the pertubation parameter $\mu$ is reintroduced. Notice that such a truncation is consistent.

If we pick for $g$ and $\theta$ generic $n \times n$ symmetric real  matrices and construct the corresponding matrix $\M$ through (\ref{Mmatrix}), $\M$ provides a coset representative of ${\Symp(2 n) \over \U(n)}$. Upon integration of the auxiliary fields $g$ and $\theta$, this choice yields a self-dual Born-Infeld theory featuring on-shell $\U(n)$ duality, which coincides with the Lagrangian constructed in \cite{ABMZ}:
\be
\LL={2 \mu^2} \left[1-\mbox{SymTr}\sqrt{\II+{1 \over 2 \mu^2} \mathfrak{F}-{1 \over 16 \mu^4} ({}^*\mathfrak{F})^2}\right].\label{SymTrAct}
\ee
It is defined in terms of the symmetrized trace:
\be
\mbox{SymTr}(M_{rs}(\FF, {}^* \FF))=\Tr \left[\left( \begin{array}{c}
r+s\\ r\end{array}\right)^{-1}{1 \over r! s!}\left({\partial \over \partial \mu}\right)^r \left({\partial \over \partial \nu}\right)^s (\mu \FF+\nu\,  {}^* \FF)^{r+s}\right],
\ee
which for each monomial $M_{rs}(\FF, {}^* \FF)$ of degree $r$ in $\FF$ and $s$ in  ${}^* \FF$ appearing in the expansion of the square root, takes the trace of the symmetrized product, i.e. the trace of the sum of all possible permutations of monomials of degree $r$ in $\FF$ and s in ${}^* \FF$.\footnote{This operation should not be confused with the symmetrization adopted in \cite{Tse1997} over the products of gauge generators in the definition of the non-Abelian Born-Infeld Lagrangian. The prescription of \cite{Tse1997} amounts to symmetrizing over the $I,J,\dots$ indices in the adjoint of the gauge group $K$ within each term in the expansion of the action proposed in the same paper.  }

Let us remark that specializing to the case $n=2$ this Lagrangian does not coincide with the Lagrangian with $\U(2)$ duality found in \cite{FAY2016}, as can be inferred immediately e.g. from the fact that it does not contain terms with square roots inside the square root.

\section{2-field Born-Infeld Theories with $\bm{\U(1)}$, $\bm{\U(1) \times \U(1)}$ and $\bm{\SU(2)}$ symmetry}
\label{2field}

In this Section we study various examples where it is explicitly possible to integrate the fields. We are able to recover various non-linear Born-Infeld like theories that have been described in \cite{FAY2015,FAY2016}, thus explaining the particular form of the corresponding Lagrangians and their symmetry properties.

\subsection{Electric $\bm{\U(1)}$}\label{u1sol}

The first example is the case of a Lagrangian with manifest (electric) $\U(1)$ symmetry. To this aim we start with the coset representative of ${\SL(2) \over \U(1)}$ in the spin-${1 \over 2}$ representation, defined in the solvable Iwasawa decomposition by:
\be
L=\left(
\begin{array}{cc}
1 &0\\
-y &1
\end{array}
\right)
\left(\begin{array}{cc}
e^{-\phi/2} &0\\
0&e^{\phi/2}
\end{array}
\right)
=
\left(\begin{array}{cc}
e^{-\phi/2} &0\\
-y e^{-\phi/2}&e^{\phi/2}
\end{array}
\right)\,,\label{solL}
\ee
where $\phi$ corresponds to the Cartan subalgebra and $y$ to the nilpotent generator.

Then we consider its diagonal embedding in ${\Symp(4) \over \U(2)}$ given by:
\be
\bf{L}=\left(\begin{array}{cc}
L&\OO\\
\OO&(L^T)^{-1}
\end{array}
\right)
\label{LmatrixU1}
\ee
and use the matrix $\bf{L}$ to construct the symmetric symplectic matrix:
\be
\M=\bf{L} \, \bf{L}^T=\left(\begin{array}{cccc}
 {1 \over x}& -{y \over x} & 0& 0\\
 -{y \over x} & \frac{x^2 + y^2}{x}& 0& 0\\
 0& 0& \frac{x^2 + y^2}{x}& {y \over x}\\
 0&0& {y \over x}& {1\over x}
\end{array}
\right),
\label{MU1el}
\ee
where we have set $x=e^\phi$.

Comparing the particular expression (\ref{MU1el}) for $\mathcal{M}$ with the general formula~(\ref{Mmatrix}) in terms of $g$ and $\theta$ yields:
\be
g=\left(\begin{array}{cc}
\frac{1}{x} & -\frac{y}{x}\\
 -\frac{y}{x}& \frac{x^2 + y^2}{x}
\end{array}
\right); \quad
\theta=
\left(\begin{array}{cc}
0&0\\
0&0
\end{array}
\right).
\ee
We insert these expressions into the linear Lagrangian (\ref{linbi}):
\be
\mathcal{L}= -\frac{1}{4}\, {F}^T_{ {\mu} {\nu}}\,g\, {F}^{ {\mu} {\nu}}-\frac{\mu^2}{2}\,\Tr(\mathcal{ M})+2\mu^2=-\frac{1}{4}\, {F}^T_{ {\mu} {\nu}}\,g\, {F}^{ {\mu} {\nu}}-\frac{\mu^2}{2}\,\Tr(g+g^{-1})+2 \mu^2
\label{U1ellinbi}
\ee
and find the equations of motion for the auxiliary fields $g$ and $\theta$, which are parametrized by $x$ and $y$:
\bea
{\partial \LL \over \partial x}&=&{1 \over \mu^2}  F^1_{\mu \nu} {F^1}^{\mu \nu}-x^2 \left({1 \over \mu^2}  F^2_{\mu \nu} {F^2}^{\mu \nu}+4\right)+y^2 \left({1 \over \mu^2}  F^2_{\mu \nu} {F^2}^{\mu \nu}+4\right)-{2 \over \mu^2}  y F^1_{\mu \nu} {F^2}^{\mu \nu}+4=
0; \nn \\
{\partial \LL \over \partial y}&=&{1 \over \mu^2}  F^1_{\mu \nu} {F^2}^{\mu \nu}-y \left({1 \over \mu^2}  F^2_{\mu \nu} {F^2}^{\mu \nu}+4\right)=0.
\label{eomU1}
\eea
Eqns. (\ref{eomU1}) are polynomial of degree 2 in $x$ and $y$ and can be solved, allowing to integrate out the auxiliary fields $g$ and $\theta$:
\bea
x&=&\frac{\sqrt{F^1_{\mu \nu} {F^1}^{\mu \nu} F^2_{\mu \nu} {F^2}^{\mu \nu}-(F^1_{\mu \nu} {F^2}^{\mu \nu})^2+4\mu^2 (F^1_{\mu \nu} {F^1}^{\mu \nu}+F^2_{\mu \nu} {F^2}^{\mu \nu})+16\mu^2}}{F^2_{\mu \nu} {F^2}^{\mu \nu}+4 \mu^2};\\
y&=&\frac{F^1_{\mu \nu} {F^2}^{\mu \nu}}{F^2_{\mu \nu} {F^2}^{\mu \nu}+4 \mu^2}.
\eea
In this way we recover the non-linear Lagrangian~\cite{FAY2015}:
\be
\mathcal{L}=2 \mu^2\left[1-\sqrt{1+\frac{1}{4\mu^2} \left(F^1_{\mu \nu} {F^1}^{\mu \nu}+F^2_{\mu \nu} {F^2}^{\mu \nu}\right)+
\frac{1}{16 \mu^4} \left( F^1_{\mu \nu} {F^1}^{\mu \nu} \, F^2_{\rho \sigma} {F^2}^{\rho \sigma}-(F^1_{\mu \nu} {F^2}^{\mu \nu})^2\right)}\,\right].
\label{LagU1el}
\ee
Since $\theta$ vanishes and the embedding is diagonal, this Lagrangian has manifest (electric) $\U(1)$ duality. It is doubly self-dual under a Legendre tranform in both vectors.

Notice that the same Lagrangian can be obtained by starting from a generic symmetric matrix $g=\left(\begin{array}{cc}
x & y\\
 y& z
\end{array}
\right)$
and adding an extra auxiliary field $s$ as Lagrange multiplier in the Lagrangian (\ref{U1ellinbi}) to implement the constraint of unit determinant in order to restrict to the special linear group:
\bea
\LL&=&-\frac{1}{4}\, {F}^T_{ {\mu} {\nu}}\,g\, {F}^{ {\mu} {\nu}}-\frac{\mu^2}{2}\,\Tr(g+g^{-1})+2\mu^2+s (\mbox{Det}(g)-1)
\label{Lagdet}\\
&=&
\frac{1}{4} \left(-x F^1_{\mu \nu} {F^1}^{\mu \nu}-2 y F^1_{\mu \nu} {F^2}^{\mu \nu}-z F^2_{\mu \nu} {F^2}^{\mu \nu}-4 s \left(-x z+y^2+1\right)\right.\nn\\
&&\left. +\frac{2\mu^2 (x+z) \left(x z-y^2+1\right)}{\left(y^2-x z\right)}\right)+2 \mu^2.
\nn
\eea
We compute the equations of motion for the auxiliary fields parametrized by $x,y,z$ and $s$:
\bea
{\partial L \over \partial x}&=&
\frac{4 s z \left(y^2-x z\right)^2+2\mu^2\left(-x^2 z^2+2 x y^2 z-y^4+y^2+z^2\right)-F^1_{\mu \nu} {F^1}^{\mu \nu} \left(y^2-x z\right)^2}{4  \left(y^2-x z\right)^2}=0;\nn\\
{\partial L \over \partial y}&=&-\frac{1}{2} F^1_{\mu \nu} {F^2}^{\mu \nu}-2 s y-\frac{\mu^2 y (x+z)}{\left(y^2-x z\right)^2}=0;
\label{eomdet}\\
{\partial L \over \partial z}&=&\frac{4 x s \left(y^2-x z\right)^2+2\mu^2\left(-x^2 z^2+x^2+2x y^2 z-y^4+y^2\right)-F^2_{\mu \nu} {F^2}^{\mu \nu} \left(y^2-x z\right)^2}{4\left(y^2-x z\right)^2}=0;\nn\\
{\partial L \over \partial s}&=&x z-y^2-1=0. \nn
\eea
These equations are solved by:
\bea
x&=&\frac{F^2_{\mu \nu} {F^2}^{\mu \nu}+4\mu^2}{\sqrt{F^1_{\mu \nu} {F^1}^{\mu \nu} F^2_{\mu \nu} {F^2}^{\mu \nu}-(F^1_{\mu \nu} {F^2}^{\mu \nu})^2+4 \mu^2  (F^1_{\mu \nu} {F^1}^{\mu \nu}+F^2_{\mu \nu} {F^2}^{\mu \nu})+16\mu^4}};\nn\\
y&=& -\frac{F^1_{\mu \nu} {F^2}^{\mu \nu}}{\sqrt{F^1_{\mu \nu} {F^1}^{\mu \nu} F^2_{\mu \nu} {F^2}^{\mu \nu}-(F^1_{\mu \nu} {F^2}^{\mu \nu})^2+4 \mu^2  (F^1_{\mu \nu} {F^1}^{\mu \nu}+F^2_{\mu \nu} {F^2}^{\mu \nu})+16 \mu^4}};\nn\\
z&=&\frac{F^1_{\mu \nu} {F^1}^{\mu \nu}+4\mu^2}{\sqrt{F^1_{\mu \nu} {F^1}^{\mu \nu} F^2_{\mu \nu} {F^2}^{\mu \nu}-(F^1_{\mu \nu} {F^2}^{\mu \nu})^2+4 \mu^2  (F^1_{\mu \nu} {F^1}^{\mu \nu}+F^2_{\mu \nu} {F^2}^{\mu \nu})+16\mu^4}};\\
s&=& \frac{-(F^1_{\mu \nu} {F^2}^{\mu \nu})^2+F^1_{\mu \nu} {F^1}^{\mu \nu} (F^2_{\mu \nu} {F^2}^{\mu \nu}+2 \mu^2)+2 \mu^2 F^2_{\mu \nu} {F^2}^{\mu \nu}}{4 \sqrt{F^1_{\mu \nu} {F^1}^{\mu \nu} F^2_{\mu \nu} {F^2}^{\mu \nu}-(F^1_{\mu \nu} {F^2}^{\mu \nu})^2+4 \mu^2  (F^1_{\mu \nu} {F^1}^{\mu \nu}+F^2_{\mu \nu} {F^2}^{\mu \nu})+16 \mu^4}}. \nn
\eea
Substituting these expressions in the Lagrangian (\ref{Lagdet}) reproduces (\ref{LagU1el}). It shows that it is the last constraint in (\ref{eomdet}), the one enforcing the condition of unit determinant, which causes the term in $\mu^4$ to appear in the square root.
\medskip

\subsection{Magnetic $\bm{\U(1)$}}

The second example is obtained by performing a duality rotation:
\be
R= \left(
\begin{array}{cccc}
    0& 0& -1& 0\\
    0& 1& 0& 0\\
    1& 0& 0& 0\\
    0 &0& 0& 1
\end{array}
\right)
\ee
of the $\U(1)$ subgroup, which correspondingly transforms $\M$ in (\ref{MU1el})  to:
\be
\M \rightarrow  R \M R^T=
\left(
\begin{array}{cccc}
 \frac{x^2 + y^2}{x}& 0& 0& -\frac{y}{x}\\
 0& \frac{x^2 + y^2}{x}& -\frac{y}{x}& 0\\
 0 & -\frac{y}{x} &\frac{1}{x} &0\\
 -\frac{y}{x}& 0& 0& \frac{1}{x}
\end{array}
\right).
\label{MU1em}
\ee
As in the previous example, a  comparison of (\ref{MU1em}) with (\ref{Mmatrix}) gives the explicit values for $g$ and $\theta$, which can then be substituted in the linear Lagrangian (\ref{elinbi}). The last step is to eliminate the auxiliary fields by means of the equations of motion, and, with the choice of the constant as $2\mu^2$, this leads to the non-linear Lagrangian~\cite{FAY2015}:
\be
\mathcal{L}=2 \mu^2 \left[1-\sqrt{1+\frac{1}{4\mu^2} \left(F^1_{\mu \nu} {F^1}^{\mu \nu}+F^2_{\mu \nu} {F^2}^{\mu \nu}\right)-
\frac{1}{16\mu^4} \left(F^1_{\mu \nu} {}^* F{}^2{}^{\mu \nu}\right)^2}\, \right].
\label{LagU1em}
\ee
The same Lagrangian can be obtained by performing a Legendre transform on one field, i.e. adding to the Lagrangian (\ref{LagU1el}) a term
$-{1 \over 4} \varepsilon^{\mu \nu \lambda \delta} F^2_{\mu \nu} G^2_{\lambda \delta},$
expressing $F^2_{\mu \nu}$ in terms of $G^2_{\mu \nu}$ by means of its equations of motion, and then renaming the field $G^2_{\mu \nu} \rightarrow F^2_{\mu \nu}$, ${}^* G{}^2{}^{\mu \nu} \rightarrow {}^* F{}^2{}^{\mu \nu}$.

The $\U(1)$ subgroup is not embedded diagonally any more and, therefore, the $\U(1)$ duality becomes electric-magnetic and holds only on-shell on the equations of motion.

For the consistent truncation $F^1_{\mu \nu}=F^2_{\mu \nu}$ the Lagrangian~(\ref{LagU1em}) reduces to the Born-Infeld theory for one field (\ref{1fieldBorn-Infeld}).

\subsection{Magnetic $\bm{\U(1) \times \U(1)}$}

Next, we study two examples of Lagrangians which feature $\U(1) \times \U(1)$ duality. To this aim let us consider the embedding of
\be
{\SL(2) \over \U(1)} \times {\SL(2) \over \U(1)} \hookrightarrow {\Symp(4) \over \U(2)}
\ee
defined by:
\be
\M=\left(
\begin{array}{cccc}
 \frac{1}{x} & 0 & -\frac{y}{x} & 0 \\
 0 & \frac{1}{x_2} & 0 & -\frac{y_2}{x_2} \\
 -\frac{y}{x} & 0 & \frac{x^2+y^2}{x} & 0 \\
 0 & -\frac{y_2}{x_2} & 0 & \frac{x_2^2+y_2^2}{x_2} \\
\end{array}
\right),
\ee
where $x,y$ parametrize the first $\SL(2) \over \U(1)$ and $x_2,y_2$ the second. It is worth remarking that in this case both $\U(1)$ subgroups are magnetic. Integrating out the auxiliary fields $x,y,x_2,y_2$, the equations of motion factorize: the equations for $x,y$ allow to solve for $x,y$ only in terms of $F^1, {}^* F^1$, while the ones in $x_2,y_2$ allow to solve for $x_2,y_2$ solely in terms of $F^2, {}^* F^2$. Therefore, the resulting non-linear Lagrangian reduces to the sum of two one-field Born-Infeld theories (\ref{1fieldBorn-Infeld}):
\be
\LL=
\mu^2 \left[2-\sqrt{1+\frac{1}{2\mu^2} F^1_{\mu \nu} {F^1}^{\mu \nu}-
\frac{1}{16\mu^4} \left(F^1_{\mu \nu} {}^* F^1 {}^{\mu \nu}\right)^2}\, -\sqrt{1+\frac{1}{2\mu^2} F^2_{\mu \nu} {F^2}^{\mu \nu}-
\frac{1}{16\mu^4} \left(F^2_{\mu \nu} {}^* F^2 {}^{\mu \nu}\right)^2}\, \right].
\label{2fieldBorn-Infeld}
\ee

\subsection{Electric $\bm{\U(1) \, \, \times}$ magnetic $\bm{\U(1)}$}\label{U1U1}

On the other hand, if we start from a matrix $\M$ defined by:
\be
\M=\left(
\begin{array}{cccc}
 \frac{1}{x} & 0 & -\frac{y}{x} & 0 \\
 0 & \frac{1}{x} & 0 & -\frac{y}{x} \\
 -\frac{y}{x} & 0 & \frac{x^2+y^2}{x} & 0 \\
 0 & -\frac{y}{x} & 0 & \frac{x^2+y^2}{x} \\
\end{array}
\right),
\label{MU1xU1}
\ee
then, integrating out $x,y$ yields the Lagrangian:
\be
\mathcal{L}=2 \mu^2 \left[1-\sqrt{1+\frac{1}{4\mu^2} \left(F^1_{\mu \nu} {F^1}^{\mu \nu}+F^2_{\mu \nu} {F^2}^{\mu \nu}\right)-
\frac{1}{64\mu^4} \left(F^1_{\mu \nu} {}^* F{}^1{}^{\mu \nu}+F^2_{\mu \nu} {}^* F{}^2{}^{\mu \nu}\right)^2}\, \right].
\label{LagU1xU1}
\ee
The matrix $\M$ describes an  $\SL(2) \over \U(1)$ transformation of ${\Symp(4) \over \U(2)}$, spanned by $x,y$. However, like the previous Lagrangian (\ref{2fieldBorn-Infeld}), the Lagrangian (\ref{LagU1xU1}) features a $\U(1)\times\U(1)$ on-shell symmetry. It coincides with the Lagrangian described  in Eq.~(5.6) of~\cite{FAY2016}. Notice that while the first $\U(1)$, which is the one corresponding to the denominator in the coset $\SL(2) \over \U(1)$ described by $x,y$, is magnetic, the second one is electric, i.e. acts diagonally on the field strengths. It is generated by the matrix
\be
u=\left(
\begin{array}{cccc}
 0 & -a & 0 & 0 \\
 a & 0 & 0 & 0 \\
 0 & 0 & 0 & -a \\
 0 & 0 & a & 0 \\
\end{array}
\right),
\ee
which commutes with the matrix $\M$ given in~(\ref{MU1xU1}), and with the magnetic $\U(1) \subset \SL(2)$. As a consequence the Lagrangian~(\ref{LagU1xU1}) can be interpreted as generated by a potential defined by matrix $\M$ which corresponds to the embedding:
\be
{\SL(2) \over \U(1)} \times {\U(1) \over \U(1)} \hookrightarrow {\Symp(4) \over \U(2)}.
\ee
Notice that when expressed in terms of one complex field strength $F=F^1+i \, F^2$, its complex conjugate $\bar F=F^1-i F^2$ and their duals, the Lagrangian (\ref{LagU1xU1}) coincides with the case $n=1$ of the complex Lagrangian constructed in~\cite{ABMZ}. Indeed, it holds that $\U(1) \times \U(1)$ can be obtained also as a subgroup of $\U(1,1) \subset \Symp(4)$, where $\U(1,1)$ describes the duality of one complex field. As a special case of the models considered in~\cite{ABMZ}, it can be supersymmetrized. One can indeed show that it satisfies the general conditions given in~\cite{FAY2016} for the existence of a supersymmetric extension.

\subsection{Electromagnetic $\bm{\SU(2)}$}

Now, we proceed with the study the embedding $\frac{\SO(1,3)}{\SO(3)}\hookrightarrow {\Symp(4) \over \U(2)}$ defined by:
\be
\M=\left(\begin{array}{cccc}
 {1 \over w} + 16 w (x^2 + y^2)& 0& 4 w y& 4 w x\\
 0& {1 \over w} + 16 w (x^2 + y^2) & 4 w x& -4 w y\\
 4 w y& 4 w x& w& 0\\
 4 w x& -4 w y& 0& w\\
\end{array}
\right).
\label{MSU2}
\ee
It is obtained by starting from the representation of the Lorentz group given by:
\be
\gamma_{ij}=\gamma_i-\gamma_j, \; j=0,\ldots, 3;
\ee
with the gamma matrices:
\be
\begin{array}{cc}
\gamma_0=-i \left(
\begin{array}{cccc}
 0 & 0 & 0 & -1 \\
 0 & 0 & 1 & 0 \\
 0 & -1 & 0 & 0 \\
 1 & 0 & 0 & 0 \\
\end{array}
\right); &
\gamma_1=i \left(
\begin{array}{cccc}
 1 & 0 & 0 & 0 \\
 0 & -1 & 0 & 0 \\
 0 & 0 & 1 & 0 \\
 0 & 0 & 0 & -1 \\
\end{array}
\right); \\
\gamma_2=i \left(
\begin{array}{cccc}
 0 & 0 & 0 & 1 \\
 0 & 0 & -1 & 0 \\
 0 & -1 & 0 & 0 \\
 1 & 0 & 0 & 0 \\
\end{array}
\right); &
\gamma_3=-i \left(
\begin{array}{cccc}
 0 & 1 & 0 & 0 \\
 1 & 0 & 0 & 0 \\
 0 & 0 & 0 & 1 \\

 0 & 0 & 1 & 0 \\
\end{array}
\right),
\end{array}
\ee
and subsequently by constructing the Iwasawa decomposition of $\frac{\SO(1,3)}{\SO(3)}$:
\be
{\bf L}=e^{x N_1+y N_2} e^{\phi h},
\label{LSU2}
\ee
where the nilpotent generators are chosen as
$
N_1=\gamma_{01}+\gamma_{21}$; $N_2=\gamma_{03}+\gamma_{23},
$
and the Cartan element as $h=\gamma_{02}$. Since the matrix {\bf L} in (\ref{LSU2}) is symplectic, the symmetric symplectic coset representative (\ref{MSU2}) is then constructed as $\M=\bf{L} \, \bf{L}^T$, where we have set $w=e^{4 \phi}$.

With the same procedure as in the previous examples, we can now use (\ref{Mmatrix}) to compute $g$ and $\theta$ and then plug the expressions in the linear Lagrangian (\ref{linbi}). Finally, we can integrate out the auxiliary fields by means of their equations of motion, and fixing the constant as $2 \mu^2$, this allows us to recover the Lagrangian~\cite{FAY2016}:
\be
\LL=
2 \mu^2 \left[1-\sqrt{1+\frac{1}{4\mu^2} \left(F^1_{\mu \nu} {F^1}^{\mu \nu}+F^2_{\mu \nu} {F^2}^{\mu \nu}\right)-\frac{1}{64\mu^4} A}\, \right],
\ee
where
\be
A=\left(F^1_{\mu \nu} {}^* F{}^1{}^{\mu \nu}\right)^2+\left(F^2_{\mu \nu} {}^* F{}^2{}^{\mu \nu}\right)^2+4 \left(F^1_{\mu \nu} {}^* F{}^2{}^{\mu \nu}\right)^2-2 F^1_{\mu \nu}{}^* {F^1}^{\mu \nu} \, F^2_{\alpha \beta} {}^*{F^2}^{\alpha \beta}.
\ee
This theory has $\SU(2)$ duality symmetry.


\section{$\bm{n}$-field Born theory with $\bm{\SO(n)}$ symmetry}
\label{SOn}

In this Section we obtain a new theory with $\SO(n)$ duality symmetry. It is particularly interesting, because it can be generalized to non-Abelian field-strenghts: $F_{\mu \nu}=\partial_\mu A_\nu-\partial_\nu A_\mu-g_c [A_\mu,A_\nu]$. It is a multi-field generalization of Born theory~(\ref{1fieldBorn}).

We start with the diagonal embedding ($\theta=\OO$):
\be
{\GL(n)\over \SO(n)} \hookrightarrow {\Symp(2n,\RR) \over \U(n)}
\ee
and the linear Lagrangian:
\be
\mathcal{L}=-\frac{1}{4} \Tr (\FF g)-{\mu ^2 \over 2} \Tr\left(g+g^{-1}\right)+n \mu^2.
\label{sonlinbi}
\ee
It is possible to explicitly solve the equations of motion of the auxiliary field $g$:
\be
{\delta\mathcal{L}\over \delta g}=-{1 \over 4} \FF+{\mu^2 \over 2}(g^{-2}-\II)=\OO
\ee
and we find:
\be
g=\left(\sqrt{\II +{1 \over 2\mu^2} \, \FF}\right)^{-1}\,.
\label{gSOn}
\ee
The matrix square root is defined according to the following prescription. If $A$ is a diagonalizable matrix with positive eigenvalues $a_i$, its square root $\sqrt{A}$, in the basis in which $A$ is diagonal, is defined as $\sqrt{A_D}={\rm diag}(\sqrt{a_i})>0$.

Substituting the expression (\ref{gSOn}) for $g$ in (\ref{sonlinbi}) yields the new non-linear Born-Infeld type Lagrangian:
\bea
\mathcal{L}&=&-{\mu^2\over 2} \Tr\left[\left(\II+{1 \over 2\mu^2}\FF\right) g \right]-{\mu^2 \over 2} \Tr\sqrt{\II +{1\over 2\mu^2} \, \FF}+{n \mu^2} \nn \\
&=&-\mu^2{\rm Tr}\sqrt{\II +{1\over 2\mu^2} \, \FF}+n \mu^2.
\label{LagSOn}
\eea
It has a simple form, because it does not contain the $\mu^4$ term inside the square root, due to the fact that $\GL(n)$ does not impose the constraint of unit determinant.

By construction this Lagrangian has manifest $\SO(n)$ duality.
The $\SO(n)$ subgroup is embedded diagonally.
The theory is self-dual under a Legendre transform on all the fields. To see this, we add a term $-{1 \over 4} \varepsilon^{\mu \nu \lambda \delta} F^I_{\mu \nu} G^I_{\lambda \delta}$ to the Lagrangian (\ref{LagSOn}):
\be
\LL=-\mu^2 {\rm Tr}\sqrt{\II +{1 \over 2\mu^2} \, \FF}+n \mu^2-{1 \over 4} \varepsilon^{\mu \nu \lambda \delta} F^I_{\mu \nu} G^I_{\lambda \delta},
\label{LegLagSOn}
\ee
and we eliminate $F^I_{\mu \nu}$ in favour of $G^I_{\lambda \delta}$ by means of its equations of motion. To this aim, we consider the power series expansion in ${1\over \mu^2}$ of the quantity:
\be
f=\sqrt{\II +{1\over 2\mu^2} \, \FF}=\sum_{n=0}^\infty a_n \left({1 \over 2\mu^2}\right)^n \FF^n, \quad \mbox{where } a_n=\frac{(-1)^n (2 n)!}{(1-2 n) 4^n (n!)^2}.
\ee
Then, defining $d\FF^{IJ}=2 F^{(I}_{\mu \nu} \, dF^{J)\, \mu\nu}$, we have:
\be
df=\sum_{n=1}^\infty a_n \left({1 \over 2\mu^2}\right)^n\,( d\FF \, \FF^{n-1}+\FF \, d\FF\, \FF^{n-2}+\ldots),
\ee
from which by the cyclic property of the trace it follows:
\be
\Tr(df)=\sum_{n=1}^\infty a_n \left({1 \over 2\mu^2}\right)^n n \Tr\left(\FF^{n-1} d\FF\right)=\Tr\left[{1 \over 2\mu^2} \frac{1}{2 \sqrt{\II +{1 \over 2\mu^2} \, \FF}} \, d\FF\right].
\ee
Therefore, the equations of motion for $F^I_{\mu \nu}$ amount to:
\bea
{\partial \LL \over \partial F^I_{\mu \nu}}&=&-{1 \over 2} F^{J \, \mu \nu} \left(\II +{1 \over 2 \mu^2} \, \FF\right)^{-{1 \over 2}}_{IJ}-{1 \over 4} \varepsilon^{\mu \nu \lambda \delta} G_{I \, \lambda \delta}=0 \nn \\
\Longrightarrow
F^{I \, \mu \nu}&=&-{1 \over 2} \left(\sqrt{\II +{1 \over 2 \mu^2} \, \FF}\right)^{IJ} \varepsilon^{\mu \nu \lambda \delta} G_{J \, \lambda \delta}.
\label{Feom}
\eea
We use this expression for $F^{I \, \mu \nu}$ to compute:
\be
\FF=-\sqrt{\II +{1 \over 2 \mu^2} \, \FF} \,\, \GG \,\, \sqrt{\II +{1 \over 2 \mu^2} \, \FF}, \mbox{ where } \GG=G_{\lambda \delta} \, G^{T \, \lambda \delta}.
\ee
This allows us to express $\FF$ in terms of $\GG$:
\be
\FF=-\GG\left(\II +{1 \over 2 \mu^2} \, \GG\right)^{-1} \Longrightarrow
\left(\II +{1 \over 2 \mu^2} \, \FF\right)^{1\over 2}=\left(\II +{1 \over 2 \mu^2} \, \GG\right)^{-\frac{1}{2}}.
\label{Lagterm1}
\ee
Moreover, from (\ref{Feom}) it holds:
\be
\varepsilon^{\mu \nu \lambda \delta} F^I_{\mu \nu} G^I_{\lambda \delta}=
2\, \Tr\left[ \GG \left(\II +{1 \over 2 \mu^2} \, \GG\right)^{-\frac{1}{2}}\right].
\label{Lagterm2}
\ee
Now, we can substitute (\ref{Lagterm1}) and (\ref{Lagterm2}) in the transformed Lagrangian (\ref{LegLagSOn}):
\bea
\LL&=&-\mu^2\Tr\left[\left(\II +{1 \over 2 \mu^2} \, \GG\right)^{-\frac{1}{2}}\right]-{1\over 2}\Tr\left[ \GG \left(\II +{1 \over 2 \mu^2} \, \GG\right)^{-\frac{1}{2}}\right]+n\mu^2\no
&=&-\mu^2\Tr\left[\left(\II +{1 \over 2 \mu^2} \, \GG\right)^{-\frac{1}{2}} \left(\II +{1 \over 2 \mu^2} \, \GG\right)\right]+n \mu^2\\
&=&-\mu^2\Tr\left[\sqrt{\II +{1 \over 2 \mu^2} \, \GG}\right]+n \mu^2
\nonumber
\eea
to finally explicitly verify that it has the same form of (\ref{LagSOn}) and hence that  (\ref{LagSOn}) is, indeed,  self-dual.

\section{The Euler Parametrization}
\label{euler}

There is another parametrization for the coset representative of a homogenous symmetric space besides the solvable Iwasawa decomposition considered previously. It is a generalization \cite{CC} of the Euler angles for $\SU(2)$. This choice of coordinates is particularly suitable for our analysis since it makes the symmetry properties of the Lagrangian manifest. In particular the potential $\mathcal V=\Tr(\M)$ turns out to be independent of the angular coordinates and only to depend on the properties of the Cartan subalgebra $\frak{a}$.

We pick a maximal torus in ${G \over H}$, i.e. a Cartan subalgebra with the maximal number of generators $\frak{a}=\{a_i\}$, $i=1,\ldots,r$ with $r=$rank$({G \over H})$ in $G \over H$. Then, if we denote by $\mathcal{A}$ the Abelian subgroup generated by $\frak{a}$, and by N$(\mathcal{A})$ its normalizer in $H$, the Euler parametrization can be written as:
\be
{\bf L}={H \over \mbox{N}(\mathcal{A})} \mathcal{A}.
\ee
The interesting feature of the Euler frame is that the expression for the potential ${\mathcal V}=\Tr(\M)$ simplifies significantly due to its $H$ invariance:
\bea
\Tr(\M)&=&\Tr({\bf L} \, {\bf L}^T)=\Tr\left[{H \over \mbox{N}(A)} \mathcal{A} \mathcal{A}^T \left({H \over \mbox{N}(\mathcal{A})}\right)^T\right]=\Tr(\mathcal{A} \, \mathcal{A}^T)\\
&=&\Tr\left[ e^{\sum_{i=1}^r \phi_i (a_i+a_i^T)}\right]=\Tr\left[ e^{\sum_{i=1}^r 2 \phi_i a_i} \right], \nn
\eea
since $a_i=a_i^T$ and $H\subseteq \U(n)$. The potential depends only on the Cartan generators outside of $H$, i.e. the non-compact part of the Abelian subalgebra, in an exponential form reminiscent of a Toda model. This shows that $\mathcal V$ has a number of flat directions equal to ${\rm dim}({G\over H})-{\rm rank}({G\over H})$.
In this parametrization $\mu$ enters only the equations for the $r$ scalars $\phi_i$, but not those for the angles.
Being the potential term in the Lagrangian (\ref{linbi}) independent of the angular parameters, the equations of motion (\ref{Heom}) for these scalars reduce to a subset:
\be
\mathbb{F}^T t_\alpha \, \CC \, \,{}^* \mathbb{F}=0
\label{GaZu2}
\ee
of the Gaillard-Zumino conditions (\ref{GaZu}), where $\alpha$ only parametrizes the generators of $H/N(A)$.\par
As a simple example of the Euler parametrization we can consider the ${\rm SL}(2)/{\rm SO}(2)$ manifold. The coset representative has the following form:
\be
L=\left(
\begin{array}{cc}
\cos(\psi) &\sin(\psi)\\
-\sin(\psi) &\cos(\psi)
\end{array}
\right)
\left(\begin{array}{cc}
e^{-\varphi/2} &0\\
0&e^{\varphi/2}
\end{array}
\right)\,,
\ee
where $\psi$ is the angular variable.
Computing $\mathcal{M}(\psi,\varphi)$ as $LL^T$ we can extract from it the corresponding expressions for $g$ and $\theta$:
\begin{equation}
g=\frac{1}{e^{ \varphi } \cos ^2(\psi )+e^{-\varphi }\sin ^2(\psi )}\,,\,\,\,\theta=\frac{ \sin (2 \psi ) \sinh (\varphi )}{e^{\varphi } \cos ^2(\psi )+e^{-\varphi }\sin ^2(\psi )}\,.
\end{equation}
The potential clearly only depends on $\phi$: ${\rm Tr}(\M)=e^{-\varphi }+e^{\varphi }$.
The relation between the solvable coordinates $y,\,x=e^\phi$ used in (\ref{solL}) and the Euler ones $\psi,\,\varphi$  is the following:
\begin{equation}
x=\frac{1}{e^{-\varphi }\cos ^2(\psi )+e^{\varphi } \sin ^2(\psi )}\,\,;\,\,\,\,y=-\frac{\sin (2 \psi ) \sinh (\varphi )}{e^{-\varphi } \cos ^2(\psi )+e^{\varphi } \sin ^2(\psi )}\,.
\end{equation}
The Jacobian $J=\frac{\partial {\rm (x,y)}}{\partial {\rm (e^\varphi,\,\psi)}}$ has determinant:
\begin{equation}
{\rm det}(J)=-\frac{-1+e^{2 \varphi }}{\left(\cos ^2(\psi )+e^{2 \varphi } \sin ^2(\psi )\right)^2}\,,
\end{equation}
which vanishes for $\varphi=0$. This is analogous to what happens with the Cartesian and polar coordinates on the plane: the corresponding Jacobian is singular when $r=0$. Consequently, when integrating out the auxiliary fields in the Euler parametrization, the equations have one more solution
than in the solvable parametrization, which corresponds to:
\begin{equation}
\varphi=0\,\,;\,\,\,\,\tan(2\psi)=\frac{F_{\mu\nu} F^{\mu\nu}}{F_{\mu\nu} {}^*F^{\mu\nu}}\,.
\end{equation} 
This solution separately minimizes the kinetic terms and the potential in the quadratic Lagrangian and yields the ordinary Maxwell Lagrangian:
\begin{equation}
 \mathcal{L}=-\frac{1}{4}\,F_{\mu\nu} F^{\mu\nu}-\mu^2\,.
 \end{equation}
Therefore the Euler parametrization admits, besides the non-linear model, also ordinary Maxwell theory as a solution.
\section{The Born-Infeld theory associated with the ${\bf t^3}$ model}
\label{t3}

The manifold $\M$ associated to $t^3$ model in supergravity is the spin ${3 \over 2}$ symmetric symplectic representation of the coset ${\SL(2) \over \U(1)}$:
\be
\M=
\left(
\begin{array}{cccc}
 \frac{\left(b^2+y^2\right)^3}{b^3} & -\frac{3 y \left(b^2+y^2\right)^2}{b^3} & \frac{y^3}{b^3} & \frac{y^2 \left(b^2+y^2\right)}{b^3} \\
 -\frac{3 y \left(b^2+y^2\right)^2}{b^3} & \frac{3 \left(b^4+4 y^2 b^2+3 y^4\right)}{b^3} & -\frac{3 y^2}{b^3} & -\frac{3 y^3+2 b^2 y}{b^3} \\
 \frac{y^3}{b^3} & -\frac{3 y^2}{b^3} & \frac{1}{b^3} & \frac{y}{b^3} \\
 \frac{y^2 \left(b^2+y^2\right)}{b^3} & -\frac{3 y^3+2 b^2 y}{b^3} & \frac{y}{b^3} & \frac{b^2+3 y^2}{3 b^3} \\
\end{array}
\right),
\label{Mt3}
\ee
where we have picked $d=6$ for the normalization of the potential ${d \over 6} z^3$.

This expression (\ref{Mt3}) can be obtained by starting from the Iwasawa decomposition:
\be
N_0=\left(
\begin{array}{cccc}
 0 & -\frac{1}{\sqrt{3}} & 0 & 0 \\
 0 & 0 & 0 & -2 \sqrt{3} \\
 0 & 0 & 0 & 0 \\
 0 & 0 & \frac{1}{\sqrt{3}} & 0 \\
\end{array}
\right);
\quad
h=\left(
\begin{array}{cccc}
 \frac{\sqrt{3}}{2} & 0 & 0 & 0 \\
 0 & \frac{1}{2 \sqrt{3}} & 0 & 0 \\
 0 & 0 & -\frac{\sqrt{3}}{2} & 0 \\
 0 & 0 & 0 & -\frac{1}{2 \sqrt{3}} \\
\end{array}
\right).
\ee
Here $h$ generates the Cartan subalgebra, while $N_0$ is the nilpotent element.
Then we define:
\be
{\bf L}=e^{\sqrt{3}y N_0}e^{\sqrt 3 \log(b) h}=
\left(
\begin{array}{cccc}
 b^{3/2} & -\sqrt{b} y & \frac{y^3}{b^{3/2}} & \frac{3 y^2}{\sqrt{b}} \\
 0 & \sqrt{b} & -\frac{3 y^2}{b^{3/2}} & -\frac{6 y}{\sqrt{b}} \\
 0 & 0 & \frac{1}{b^{3/2}} & 0 \\
 0 & 0 & \frac{y}{b^{3/2}} & \frac{1}{\sqrt{b}} \\
\end{array}
\right)
\ee
However, this frame is obtained by dimensional reduction from 5 dimensional supergravity. As a consequence, the parity operator is given by:
$\mathbb{P}=\rm{diag}(-1,1,1,-1)$, i.e. under $\mathbb{P}$:
\be
A^0_\mu \rightarrow -A^0_\mu; \quad
A^1_\mu \rightarrow A^1_\mu; \quad y \rightarrow -y; \quad b \rightarrow b.
\ee
Moreover, $\M$ is defined as $\M={\bf L\cdot O\cdot O\cdot L}^T$, where $
{\bf O}=\rm{diag}(1,\sqrt{3},1,\frac{1}{\sqrt{3}})$.
In other words, in this basis, which is natural when performing a dimensional reduction from 5 to 4 dimensions, it turns out that in the origin $b=1, y=0$ we have $\M(1,0)=\rm{diag}(1,3,1,{1\over 3})$, and it is this matrix $\M(1,0)$ which is left invariant by the $\U(1)$ subgroup, and not the identity matrix. As a consequence, the matrix $N$ in (\ref{linbi}) is not the identity, but it has to be fixed to $N=({\bf O\cdot O})^{-1}=\rm{diag}(1,\frac{1}{3},1,3)$.

A consistent truncation is:
\be
y=0; \quad F^1_{\mu \nu} {F^2}^{\mu \nu}=0; \quad  F^2_{\mu \nu} {}^* {F^2}^{\mu \nu}=0,
\ee
which amounts to setting:
\be
y=0; \quad \vec E_1=\vec B_2=\vec 0.
\ee
Then the Lagrangian reduces to:
\be
\mathcal{L}=-{b^3 \over 4} F^1_{\mu \nu} {F^1}^{\mu \nu}-{3 \over 4} b F^2_{\mu \nu} {F^2}^{\mu \nu}-{1 \over 2} \left(b^3+\frac{1}{b^3}+b+\frac{1}{b}\right) \mu ^2,
\ee
and the equation of motion for $b$ to:

\be
{\delta \mathcal{L} \over \delta b}={1 \over 4} \left(\left(\frac{6}{b^4}-6 b^2+\frac{2}{b^2}-2\right) \mu ^2-3 \left(b^2 F^1_{\mu \nu} {F^1}^{\mu \nu}+F^2_{\mu \nu} {F^2}^{\mu \nu}\right)\right)=0,
\ee
which is solved by:
$$
b= \frac{1}{3} \sqrt{\frac{-B^{1 \over 3}\left(3 F^2_{\mu \nu} {F^2}^{\mu \nu}+2 \mu ^2\right)+6 \mu ^2 (3 F^1_{\mu \nu} {F^1}^{\mu \nu}+2 F^2_{\mu \nu} {F^2}^{\mu \nu})+9 \left(F^2_{\mu \nu} {F^2}^{\mu \nu}\right)^2+B^{2 \over 3}+40 \mu ^4}{B^{1 \over 3} \left(F^1_{\mu \nu} {F^1}^{\mu \nu}+2 \mu ^2\right)}}
$$
with
\begin{eqnarray*}
B&=&2862 \mu ^4 F^1_{\mu \nu} {F^1}^{\mu \nu}-198 \mu ^4 F^2_{\mu \nu} {F^2}^{\mu \nu}+729 \mu ^2 \left(F^1_{\mu \nu} {F^1}^{\mu \nu}\right)^2-54 \mu ^2 \left(F^2_{\mu \nu} {F^2}^{\mu \nu}\right)^2\\
&&-81 \mu ^2 F^1_{\mu \nu} {F^1}^{\mu \nu} F^2_{\mu \nu} {F^2}^{\mu \nu}
-27 \left(F^2_{\mu \nu} {F^2}^{\mu \nu}\right)^3+2800 \mu ^6\\
&&+9 \sqrt{3} \mu  \left(F^1_{\mu \nu} {F^1}^{\mu \nu}+2 \mu ^2\right)\left(1200 \mu ^4 (7 F^1_{\mu \nu} {F^1}^{\mu \nu}-F^2_{\mu \nu} {F^2}^{\mu \nu})-162 \left(F^2_{\mu \nu} {F^2}^{\mu \nu}\right)^3\right.\\
&&\left. +9 \mu ^2 \left(243 \left(F^1_{\mu \nu} {F^1}^{\mu \nu}\right)^2-54 F^2_{\mu \nu} {F^2}^{\mu \nu} F^1_{\mu \nu} {F^1}^{\mu \nu}
-37 (F^2_{\mu \nu} {F^2}^{\mu \nu})^2\right)+8000 \mu ^6\right)^{1\over 2},
\end{eqnarray*}
yielding for the Lagrangian:
\begin{eqnarray*}
\mathcal{L}&=&\sqrt{\left(40 \mu ^4-B^{1\over 3}\left(3 F^2_{\mu \nu} {F^2}^{\mu \nu}+2 \mu ^2\right)+6 \mu ^2 (3 F^1_{\mu \nu} {F^1}^{\mu \nu}+2 F^2_{\mu \nu} {F^2}^{\mu \nu})+9 \left(F^2_{\mu \nu} {F^2}^{\mu \nu}\right)^2+B^{2 \over 3}\right)^3\over B \left(F^1_{\mu \nu} {F^1}^{\mu \nu}+2 \mu ^2\right)}\\
&&\left(6 B^{5/3} \left(3 F^2_{\mu \nu} {F^2}^{\mu \nu}+2 \mu ^2\right)+12 B^{4/3} \left(6 \mu ^2 (3 F^1_{\mu \nu} {F^1}^{\mu \nu}-2 F^2_{\mu \nu} {F^2}^{\mu \nu})-9 \left( F^2_{\mu \nu} {F^2}^{\mu \nu}\right)^2+32 \mu ^4\right)\right.\\
&&+12 B^{2/3} \left(48 \mu ^6 (27 F^1_{\mu \nu} {F^1}^{\mu \nu}-2 F^2_{\mu \nu} {F^2}^{\mu \nu})+108 \mu ^4 \left(3 \left(F^1_{\mu \nu} {F^1}^{\mu \nu}\right)^2-2 \left(F^2_{\mu \nu} {F^2}^{\mu \nu}\right)^2\right)\right.\\
&&\left.-216 \mu ^2 \left(F^2_{\mu \nu} {F^2}^{\mu \nu}\right)^3-81 \left(F^2_{\mu \nu} {F^2}^{\mu \nu}\right)^4+1280 \mu ^8\right)\\
&&+6 \sqrt[3]{B} \left(3 F^2_{\mu \nu} {F^2}^{\mu \nu}+2 \mu ^2\right)
\left(6 \mu ^2 (3 F^1_{\mu \nu} {F^1}^{\mu \nu}+2 F^2_{\mu \nu} {F^2}^{\mu \nu})+9 \left(F^2_{\mu \nu} {F^2}^{\mu \nu}\right)^2+40 \mu ^4\right)^2\\
&&+2 B \left(18 \mu ^4 (165 F^1_{\mu \nu} {F^1}^{\mu \nu}+29 F^2_{\mu \nu} {F^2}^{\mu \nu})+27 \mu ^2 \left(27 \left(F^1_{\mu \nu} {F^1}^{\mu \nu}\right)^2+3 F^2_{\mu \nu} {F^2}^{\mu \nu} F^1_{\mu \nu} {F^1}^{\mu \nu}
\right.\right.\\
&&\left.\left.+20 \left(F^2_{\mu \nu} {F^2}^{\mu \nu}\right)^2\right)+270 \left(F^2_{\mu \nu} {F^2}^{\mu \nu}\right)^3+3104 \mu ^6\right)+\left(6 \mu ^2 (3 F^1_{\mu \nu} {F^1}^{\mu \nu}+2 F^2_{\mu \nu} {F^2}^{\mu \nu})\right.\\
&&\left.\left.+9 \left(F^2_{\mu \nu} {F^2}^{\mu \nu}\right)^2+40 \mu ^4\right)^3+B^2\right).
\end{eqnarray*}

\section*{Discussion}

We have discussed an auxiliary field description of the multi-field generalizations of both the Born and Born-Infeld theories. Our guiding principle in the construction of these non-linear models was the self-duality of their Lagrangians, which is guaranteed by the initial auxiliary field description.

Non-Abelian extensions of these models were also discussed.

It would be interesting to couple these models to matter fields and study their solitonic solutions, such as monopole solutions.
 For instance one could couple the non-Abelian Born or Born-Infeld theory with gauge symmetry $K$ to scalar fields $\Phi^I(x)$ (not to be mistaken for the auxiliary fields $\phi^s$) in the adjoint representation of $K$ and write the following Lagrangian:
 \begin{equation}
 \mathcal{L}=\mathcal{L}_{NL}[F]-\frac{1}{2}\,\kappa_{IJ}\mathcal{D}_\mu\Phi^I\mathcal{D}^\mu\Phi^J-V(\Phi)\,,
 \end{equation}
 where $\mathcal{L}_{NL}[F]$ is the non-linear action describing the non-Abelian field strengths $F^I$, $\mathcal{D}_\mu\Phi^I\equiv \partial_\mu\Phi^I-g_c\,f_{JK}{}^I\,A^J_\mu\Phi^K$ is the $K$-covariant derivative of $\Phi^I$, $\kappa_{IJ}$ the negative definite Cartan-Killing matrix of $K$ and $V(\Phi)$ is a $K$-invariant scalar potential.
As an example one could consider Born theory with gauge group $\SO(3)$ with a triplet of scalar fields $\Phi^I$ in its adjoint representation, minimally coupled to the vectors and a scalar potential of the form $V=\lambda\,(\Phi^I\Phi^I-v^2)^2$. Its monopole solutions would represent a generalization of the 't Hooft Polyakov monopole~\cite{tHP}. Instead of the non-Abelian Born theory, for whose action we could derive the compact expression(\ref{LagSOn}), one can also consider for $\mathcal{L}_{NL}[F]$ the non-Abelian generalization of the Lagrangian (\ref{SymTrAct}) of \cite{ABMZ} with $n=3$ vector multiplets transforming in the adjoint representation of the off-shell symmetry group ${\rm SO}(3)$, to be promoted to gauge group. \par
In analogy with the effective D-brane actions, it is interesting to consider non-linear models in which the scalar fields minimally coupled to the vectors and transforming in the adjoint of the gauge group $K$, enter the square root in the expression of the Lagrangian.\footnote{We are grateful to T. Ortin for suggesting this generalization.} This is effected by starting from a quadratic Lagrangian of the form:
\begin{equation}
\mathcal{L}=  -\frac{1}{4}\, {F}^T_{ {\mu} {\nu}}\,g\, {F}^{ {\mu} {\nu}}+
\frac{1}{4}\, {F}^T_{ {\mu} {\nu}}\,\theta\,{}^* {F}^{ {\mu} {\nu}}-\frac{\mu^2}{2}\,{\rm Tr}(\mathcal{ M}) +g_{IJ}\left(\frac{1}{2}\mathcal{D}_\mu\Phi^I\mathcal{D}^\mu\Phi^J-L^{IJ}(\Phi)\right)\,,
\label{linbiphi}
\end{equation}
where $L^{IJ}(\Phi)$ is a contravariant, rank-2 $K$-tensor depending on $\Phi^I$. Integrating out the auxiliary fields we would end up with a non-linear Lagrangian with gauge group $K$, which is obtained from the corresponding Lagrangian $\mathcal{L}_{NL}[F]$ without scalar fields, by replacing
\begin{equation}
\mathfrak{F}^{IJ}\,\rightarrow\,\,\,F^I_{\mu\nu} F^{J\,\mu\nu}-2\,\mathcal{D}_\mu\Phi^I\mathcal{D}^\nu\Phi^J+4\,L^{IJ}(\Phi)\,.
\end{equation}
For instance the Lagrangian (\ref{SymTrAct}) would become:
\bea
\mathcal{L}&=&-\mu^2{\rm Tr}\sqrt{\II +{1\over 2\mu^2} \, \left(\FF-2\,\mathcal{D}_\mu\Phi^I\mathcal{D}^\mu\Phi^J+4\,L^{IJ}(\Phi)\right)}+n \mu^2.
\label{LagSOnphi}
\eea
To lowest order in $1/\mu^2$ we would find the standard kinetic term $\frac{1}{2}\mathcal{D}_\mu\Phi^I\mathcal{D}^\mu\Phi^I$ for $\Phi^I$ and a scalar potential $V(\Phi)={\rm Tr}(L^{IJ}(\Phi))$.\par

However we emphasize that reproducing the non-Abelian D-brane action is not among our objectives in the present work and indeed another point to be clarified is the relation between the non-Abelian theories discussed here and the non Abelian Born-Infeld models proposed in \cite{Tse1997,LY} to describe the effective action of stacks of overlapping D3-branes.

Finally, it would be interesting to investigate the possibility of supersymmetric extensions of the models discussed in this work.

\section*{Acknowledgements}
We are grateful to  L. Andrianopoli for a careful reading of the manuscript and for interesting comments. We wish also to thank P. Aschieri, R. D'Auria, S. Ferrara, T. Ortin and A. Yeranyan for enlightening discussions.

\end{document}